\documentclass[pra,superscriptaddress,twocolumn]{revtex4-2}

\usepackage{graphicx}
\usepackage{epstopdf}
\usepackage{amsmath,amssymb,bm,mathtools} 
\usepackage{textcomp} % to have upright greek symbols eg mu
\usepackage{xspace} % for intelligent management of spaces after functions
\usepackage{soul,hyperref}

\usepackage[usenames,dvipsnames]{color}
\usepackage{cleveref}

\usepackage[caption=false]{subfig}
\usepackage{pstool}
\usepackage[caption=false]{subfig}
\usepackage{siunitx,physics}

\newcommand{\iso}[1]{\ensuremath{{^{#1}}}}
\newcommand{\tplus}{\ensuremath{^{3+}}\xspace}
\newcommand{\state}[3]{\ensuremath{{}^{#1}\! #2 _{#3}}\xspace}
\newcommand{\yso}{Y$_2$SiO$_5$\xspace}

\begin{document}
\title{Optimising the Efficiency of a Quantum Memory based on Rephased Amplified Spontaneous Emission}% Force line breaks with \\

\author{Charlotte K. Duda}
\author{Kate R. Ferguson}
\author{Rose L. Ahlefeldt}
\author{Morgan P. Hedges}
\author{Matthew J. Sellars}%

\affiliation{%
 Centre for Quantum Computation and Communication Technology, Research School of Physics, The Australian National University, Canberra ACT, Australia 
}%

\date{\today}

\begin{abstract}

We studied the recall efficiency as a function of optical depth of rephased amplified spontaneous emission (RASE), a protocol for generating entangled light. The experiments were performed on the $\state{3}{H}{4} \rightarrow \state{1}{D}{2}$ transition in the rare-earth doped crystal Pr$^{3+}$:Y$_{2}$SiO$_{5}$, using a four-level echo sequence between four hyperfine levels to rephase the emission. Rephased emission was observed for optical depths in the range of $\alpha L$ = 0.8 to 2.0 with a maximum rephasing efficiency of 14 \% observed while incorporating spin storage. This efficiency is a significant improvement over the previously reported non-classical result but is well short of the predicted efficiency. We discuss the possible mechanisms limiting the protocol's performance, and suggest ways to overcome these limits. 

\end{abstract}

\maketitle
\section{\label{sec:Intro}Introduction}
 
Quantum entangled states of light are a fundamental resource for quantum communication \cite{Kimble2008}.  These states are most commonly created with spontaneous parametric down conversion (SPDC) \cite{Burnham1970}, however, for quantum repeater applications, the light source needs to be interfaced with a quantum memory, which can be challenging for SPDC sources \cite{Fekete2013}. Alternative approaches, such as the well-known  Duan-Lukin-Cirac-Zoller (DLCZ) scheme \cite{Duan2001},  integrate the memory with the light source, eliminating the need to interface two different physical devices. These methods all operate by first generating entanglement between an ensemble of atoms and an optical field, and then later recalling the stored atomic state as a second optical field. 

Rephased amplified spontaneous emission (RASE) is an example of such an approach \cite{Ledingham2010}. RASE can be treated as an optical amplifier in which gain is generated by driving an ensemble of atoms into an optical excited state. The inverted ensemble produces an optical field, via amplified spontaneous emission (ASE), that is entangled with the state of the ensemble. This state can be later recalled as a second optical field (the RASE field) by applying a rephasing $\pi$ pulse to the optical transition in a process analogous to a Hahn echo \cite{Hahn1950}.  The storage portion of the protocol is enabled by the resonant nature of the RASE amplifier, in contrast to the far-off-resonant parametric amplifier of SPDC, for which the two optical fields can only be emitted simultaneously. If RASE is implemented in a material with multiple ground and excited states, further functionality is gained, such as long-term storage by using spin ground states, and the ability to recall the ASE and RASE fields at spectrally distinct frequencies \cite{Beavan2011}. 

A similar protocol to RASE is the atomic frequency comb implementation of the DLCZ scheme (AFC-DLCZ) \cite{Ottaviani2009}, which differs from RASE primarily in the rephasing method used: a periodic spectral grating rather than a $\pi$ pulse. Experimental demonstrations of RASE and AFC-DLCZ have used rare-earth crystals for their long optical and spin coherence times. RASE has been demonstrated in two-level \cite{Ledingham2012} and four-level \cite{Beavan2012,Ferguson2016} systems, including demonstration of non-classical correlations between the recalled fields \cite{Ledingham2012, Ferguson2016}, spin-state storage \cite{Beavan2012,Ferguson2016} and multimode storage \cite{Ferguson2016}. Similarly, AFC-DLCZ has shown non-classical, spin-state storage of multiple modes \cite{Laplane2017,Kutluer2017}.

Theoretical efficiencies of RASE and AFC-DLCZ can be as high as 100\% \cite{Stevenson2014,Jobez2014}, but all experimental efficiencies have been limited to roughly 3 \% \cite{Ferguson2016,Kutluer2017,Laplane2017}. To gain insight into these limited efficiencies we revisit the RASE demonstration in Pr\tplus:\yso of Ferguson et al. \cite{Ferguson2016}, characterising its operation as a function of the optical depth of the inverted ensemble.
 
\section{\label{sec:Exp}Experimental setup}
The experimental setup was identical to that of Ferguson et al. \cite{Ferguson2016}. All measurements were made on a 0.005\% Pr\tplus:\yso sample cut along the optical extinction axes to dimensions of $4\times5\times2$~mm (D$_1\times$D$_2\times$C$_2$). We drove the 605.977~nm (vac.) optical transition between the lowest crystal field levels of the \state{3}{H}{4} ground and \state{1}{D}{2} excited multiplets for Pr\tplus ions located in the C$_1$ symmetry site labelled ``site 1'' \cite{Equall1995}. The I = $\frac{5}{2}$ nuclear spin of the single Pr\tplus isotope \iso{141}Pr\tplus means both ground and excited states are split into three doubly-degenerate hyperfine states $|g_{1,2,3}\rangle$ and $|e_{1,2,3}\rangle$ in zero magnetic field, with splittings $\mathcal{O}$(10)~MHz. All optical transitions between these levels are allowed, although the relative oscillator strengths do vary by two orders of magnitude \cite{Nilsson2004}. 

\begin{figure}[t!]
    \centering
    \includegraphics[width=0.45\textwidth]{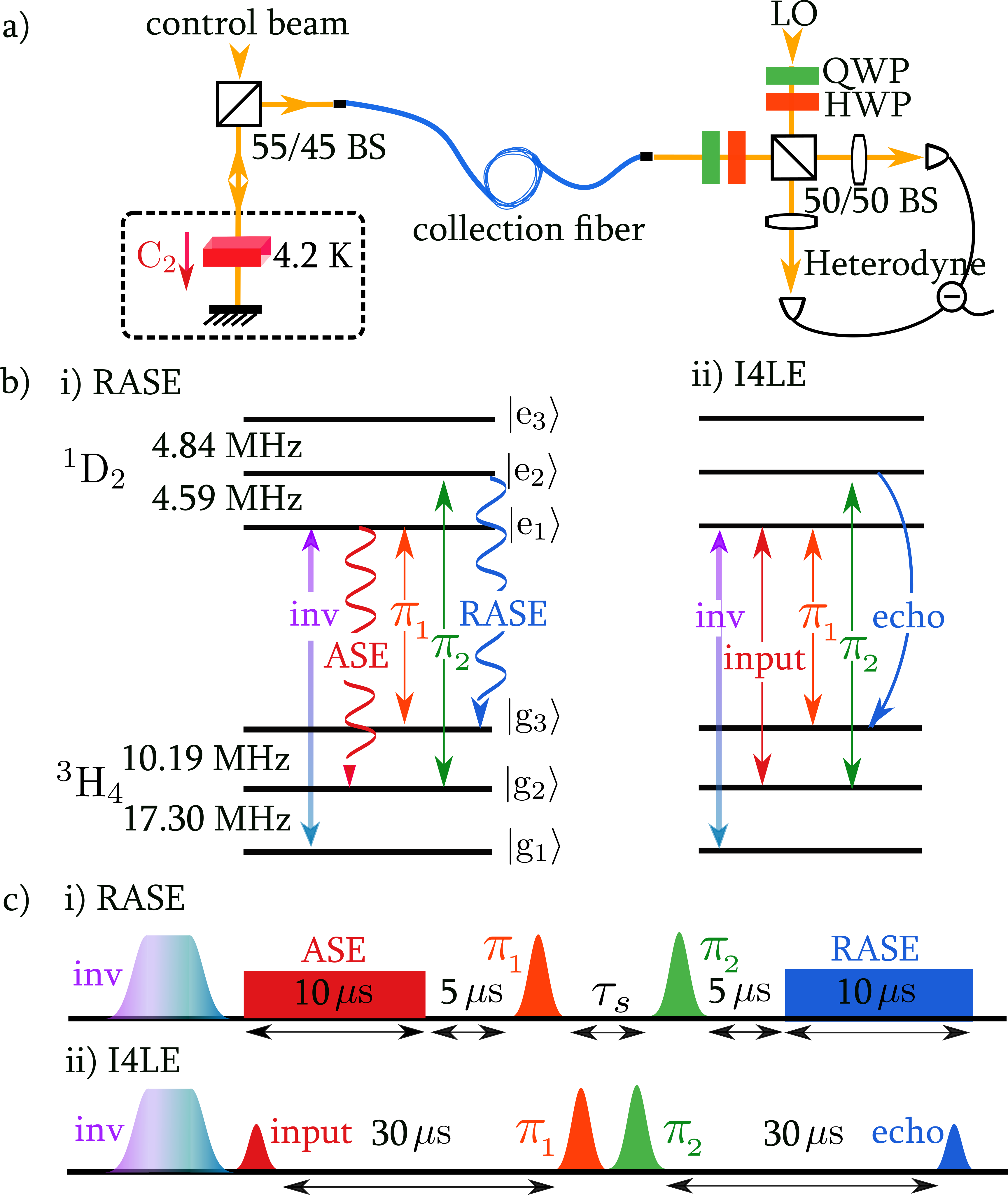}
    \caption{ (a) Experimental setup showing the control and local oscillator (LO) arms of the optical setup. HWP and QWP are half and quarter waveplates, respectively, and BS is beam splitter. (b) Hyperfine structure of the optical ground and excited states of Pr\tplus:\yso in zero field with transitions used for the RASE and inverted four-level echo (I4LE) indicated. The transitions have relative oscillator strengths of: 0.05  for $|g_{1}\rangle \leftrightarrow |e_{1}\rangle$, 0.40 for $|g_{2}\rangle \leftrightarrow |e_{1}\rangle$, 0.55 for $|g_{3}\rangle \leftrightarrow |e_{1}\rangle$, 0.60 for $|g_{2}\rangle \leftrightarrow |e_{2}\rangle$, and 0.38 for $|g_{3}\rangle \leftrightarrow |e_{2}\rangle$ \cite{Nilsson2004}.  (c) The pulse sequence used for RASE and I4LE. The rephasing pulses ($\pi_{1}$ and $\pi_{2}$) were \SI{1.7}{\micro\second} and \SI{2.5}{\micro\second} long, respectively.}
    \label{fig:exp}
\end{figure}

The experimental setup is shown in Fig. \ref{fig:exp} (a). The sample was maintained at 4.2~K in a liquid helium bucket cryostat using exchange gas cooling. Light from a frequency-stabilized Coherent dye laser was separated into a control beam and a local oscillator. The control beam was gated with a double-pass acousto-optic modulator and entered the cryostat from a window at the top, travelling down the vertical axis. The beam was reflected from a mirror placed behind the crystal, executing a double-pass of the sample, and was coupled into the collection fiber with an efficiency of 45 \%. The light propagated down the 2~mm C$_2$ axis, with a beam diameter at the sample of 47 $\pm$ \SI{20}{\micro\meter}, and was polarised along the D$_2$ axis. Signals were detected using a heterodyne detection system with a bandwidth of 20~MHz and a visibility of 90 \%. The polarisation of the local oscillator was matched to that of the signal using two waveplates. The phase of the interferometer was not locked and drifted between shots. This drift was corrected in post-processing with the use of phase-reference pulses, as described in Ref. \cite{Ferguson2016}.

Experiments began with the preparation of a narrow spectral feature from the inhomogeneously broadened line using spectral holeburning, similar to  Refs. \cite{Beavan2011,Nilsson2004}. First, we burnt a 2~MHz wide trench in the line, and then we pumped a 200~kHz wide feature composed of ions in a single hyperfine ground state back into the middle of this trench. This pumping was performed before each repetition of the experiments, limiting the repetition rate to 10~Hz. 

\begin{figure}[t!]
    \centering
    \includegraphics[width=0.45\textwidth]{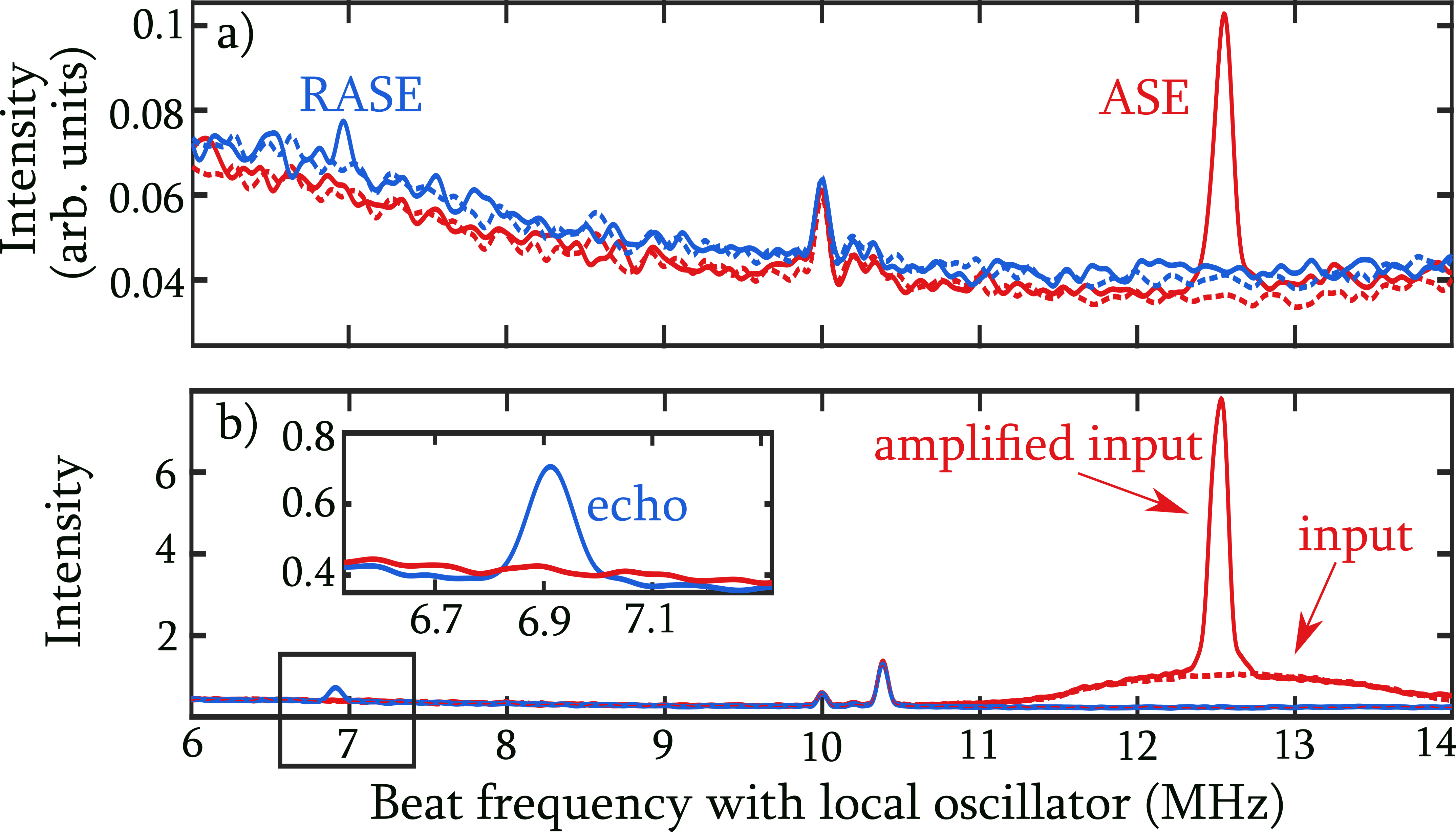}
    \caption{Spectrum of the amplified (pink) and rephased (blue) signals are shown with the respective background levels (dashed) at an optical depth of 1.4. (a) The RASE experiment, where the prepared feature amplifies the vacuum state. (b) The I4LE experiment, where the spectrum has been normalised to the input pulse.} 
    \label{fig:spec}
\end{figure}

Two pulse sequences were used to characterise the RASE protocol, shown in Fig. \ref{fig:exp} (b) and (c). The first is the RASE sequence itself, Fig. \ref{fig:exp} (b)(i) and (c)(i). A portion of the prepared ensemble was inverted with a pulse on the $\ket{g_1}\rightarrow\ket{e_1}$ transition chirped across 300~kHz in \SI{8}{\micro\second}, with the degree of inversion controlled by the pulse amplitude. The resulting amplification of the vacuum state causing emission at the ASE frequency was monitored for \SI{10}{\micro\second} and then two rephasing $\pi$ pulses were applied, with the separation between these pulses $\tau_s$, determining the spin-storage time of the sequence. We used two storage times for the results presented here, $\tau_s$ = \SI{0} and  \SI{5}{\micro\second}.  The resulting RASE signal was monitored for the same time window as the ASE, and then two phase-reference pulses (not shown) were sent through the system as described above. The second sequence, the  inverted four-level echo (I4LE) (Fig. \ref{fig:exp} (b)(ii) and (c)(ii)),  was very similar to RASE except that a $\SI{1}{\micro\second}$ long pulse replaced the vacuum state as the input to the inverted system. Importantly, this means the RASE and I4LE have different transverse  spatial modes of the stimulating field, with the weak coherent pulse mode of the I4LE defined by the input optics and the vacuum state of RASE not constrained. Comparing the behaviour of the I4LE and RASE as the amount of inversion is increased allows the sensitivity of the rephasing sequence to the input mode to be investigated. Fig. \ref{fig:spec} (a) ((b)) shows examples of the signals recorded for the RASE sequence (I4LE sequence) for the ASE (input) and RASE (echo) windows, compared to the corresponding background signals when no inversion was applied.

\section{\label{sec:gain}Optical depth measurements}

The optical depth of the created gain feature  was determined by measuring the transmission of a weak, \SI{10}{\micro\second} long, probe pulse both in the presence and absence of the inversion pulse. To confirm the validity of this measurement of the optical depth, the level of the variance of the ASE as a function the optical depth was also measured. These two quantities are related by: 
\begin{equation} \label{eq:ase_var_loss}
\langle\hat{o}^{2}_{A}\rangle=l(2e^{\alpha L}-1)+(1-l).
\end{equation}
\begin{figure}[t!]
    \centering
    \includegraphics[width=0.45\textwidth]{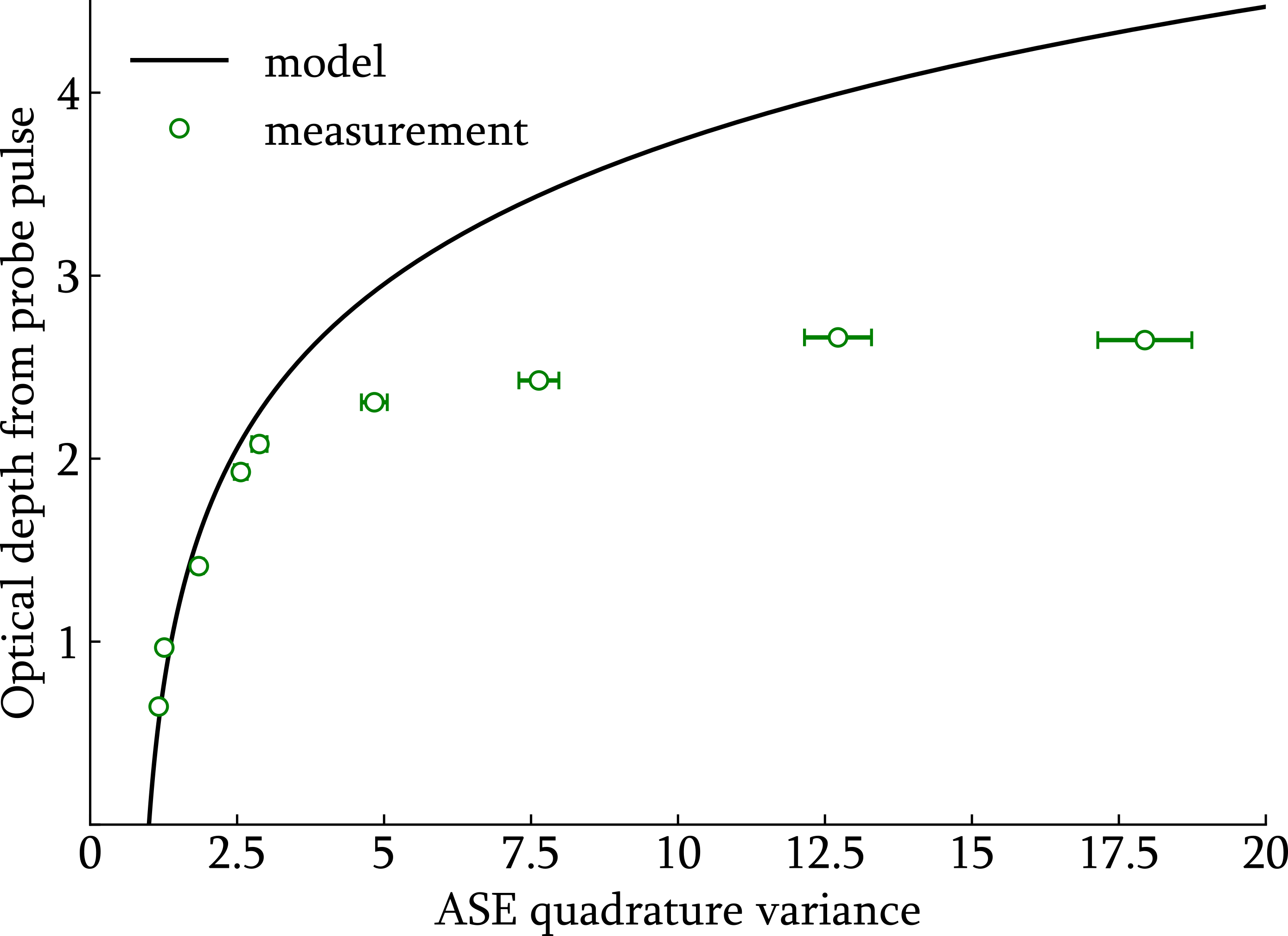}
    \caption{Optical depth measured after the inversion pulse via two measures: the optical depth seen by a probe pulse compared to the average variance of the ASE signal. 1$\sigma$ error in $x$ is shown (the $y$-error is negligible). The model was constructed using equation \ref{eq:ase_var_loss}.}
    \label{fig:ase}
\end{figure}

where $\expval{\hat{o}_{A}^{2}}$ is the average variance in the two ASE quadratures, $\alpha L$ is optical depth and $l$ represents signal losses, which are incorporated into the model as a single beamsplitter \cite{Ferguson2016}. This model assumes the light can be represented as an infinite plane wave and that the loss is independent of optical depth. For this experiment, the signal loss was 89\%, arising from the two beamsplitters and the visibility of the detection system.  

The relationship between the ASE variance (normalised to the vacuum level) and the optical depth measured is shown in Fig. \ref{fig:ase}. The optical depth was determined by averaging the signal over the full width at half maximum linewidth of the spectral feature. This linewidth increased from 100~kHz to 300~kHz as the inversion pulse intensity was increased. The ASE variance was determined for a \SI{10}{\micro\second} time window \SI{3.5}{\micro\second} after the end of the inversion pulse over a spectral range of 100~kHz. The variance predicted by equation \ref{eq:ase_var_loss} conforms to the measured variance for optical depths below 2.0. For optical depths greater than 2.0, we see the optical depth measured via the probe pulse saturates. This result indicates that probe pulse measurements with $\alpha L >$ 2.0 are likely to significantly underestimate the true value of the optical depth. For this reason, measurements of the RASE efficiency in the following section are limited to $\alpha L <$ 2.0.

The difference in the sensitivity of the ASE and optical depth measurements to the level of the inversion could indicate the presence of a loss mechanism dependent on the level of inversion. This is a consequence of the different spatial input modes, as described earlier.  Because of this difference, the probe pulse transmission measures the net optical depth: the total optical depth minus any loss in the system, while the ASE variance is sensitive to where the loss occurs along the beam path. For example, if the loss occurs at the front face of the crystal it will have no effect on the input state of light for the ASE measurement, which will continue to be the vacuum state, and hence will have no effect on the ASE variance. However, if the loss occurs on the back face of the crystal it will attenuate the output and reduce the variance. Possible loss mechanisms that will depend on the level of inversion include the off-resonant excitation by the inversion pulse populating  $|g_{2}\rangle$ and the distortion of the optical mode due to the presence of the gain feature reducing the coupling efficiency into the detection mode. Further investigation is required to determine the exact mechanism.

\section{\label{sec:eff}Rephasing efficiency measurements}

\begin{figure}[t!]
    \centering
    \includegraphics[width=0.45\textwidth]{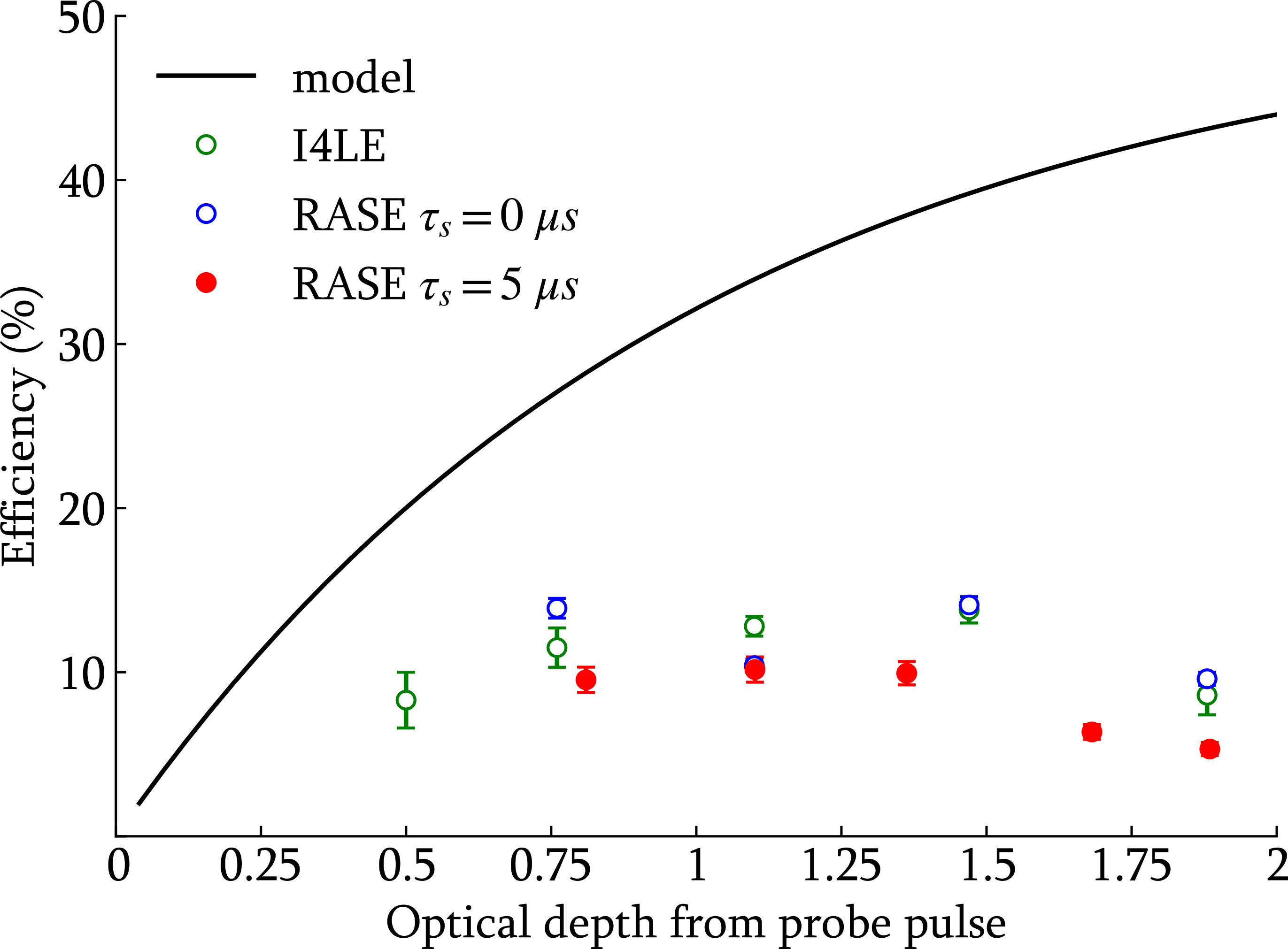}
    \caption{Efficiency of RASE and inverted four-level echo (I4LE) with optical depth. The  I4LE result was calculated using the emission spectrum while RASE result was calculated using the photon count. 1$\sigma$ error in $y$ is shown (the $x$-error is negligible). The model was constructed using equation \ref{eq:efficiency}. Both the I4LE result and model are scaled to the optical delay time used in the RASE experiment using the measured optical decay constant.}
    \label{fig:eff}
\end{figure}

Fig. \ref{fig:eff} shows the rephasing efficiencies of the RASE and I4LE protocols as a function of the measured optical depth of the initial gain feature. The I4LE efficiency was determined by taking the ratio of the amplified input pulse and the echo signal area in the emission intensity spectra over a 100~kHz range. To account for the different delay times, the I4LE result, and model curve, are scaled to the optical delay time used for the RASE experiment (\SI{20}{\micro\second}). The scaling was done using an optical decay time measured using the 4LE \cite{Beavan2011} of 59.2 $\pm$ \SI{1.4}{\micro\second}.

The RASE efficiency model curve shown in Fig. \ref{fig:eff} was constructed by taking the ratio of the mean ASE and RASE photon number expected to arrive at the detection system for a single run of the experiment, based on the theory presented by Ledingham et al. \cite{Ledingham2010}:

\begin{equation} \label{eq:efficiency}
\eta=\frac{\langle\hat{o}^{2}_{R}\rangle-1}{\langle\hat{o}^{2}_{A}\rangle-1}=\frac{1+8\sinh^{2}(\frac{\alpha L}{2})-2}{2e^{\alpha L}-2}
\end{equation}

For both RASE and the I4LE the observed efficiency as a function of optical depth follow the same trend. RASE and the I4LE efficiency peaks at roughly 14 \% in the range $\alpha L$ = 0.8 to 1.5. The observed efficiency is significantly less than that predicted by equation \ref{eq:efficiency}, with the deviation increasing with optical depth. 

That both RASE and the I4LE have the same dependency on optical depth suggests that mode distortion of the input is not a major mechanism for the reduction in efficiency with optical depth. It is likely that the decreasing efficiency seen for optical depths above 1.5 is due to the increased optical depth seen by the control pulses. 

\section{\label{sec:Insep}Inseparability Measurement}
\begin{figure}[t!]
    \centering
    \includegraphics[width=0.45\textwidth]{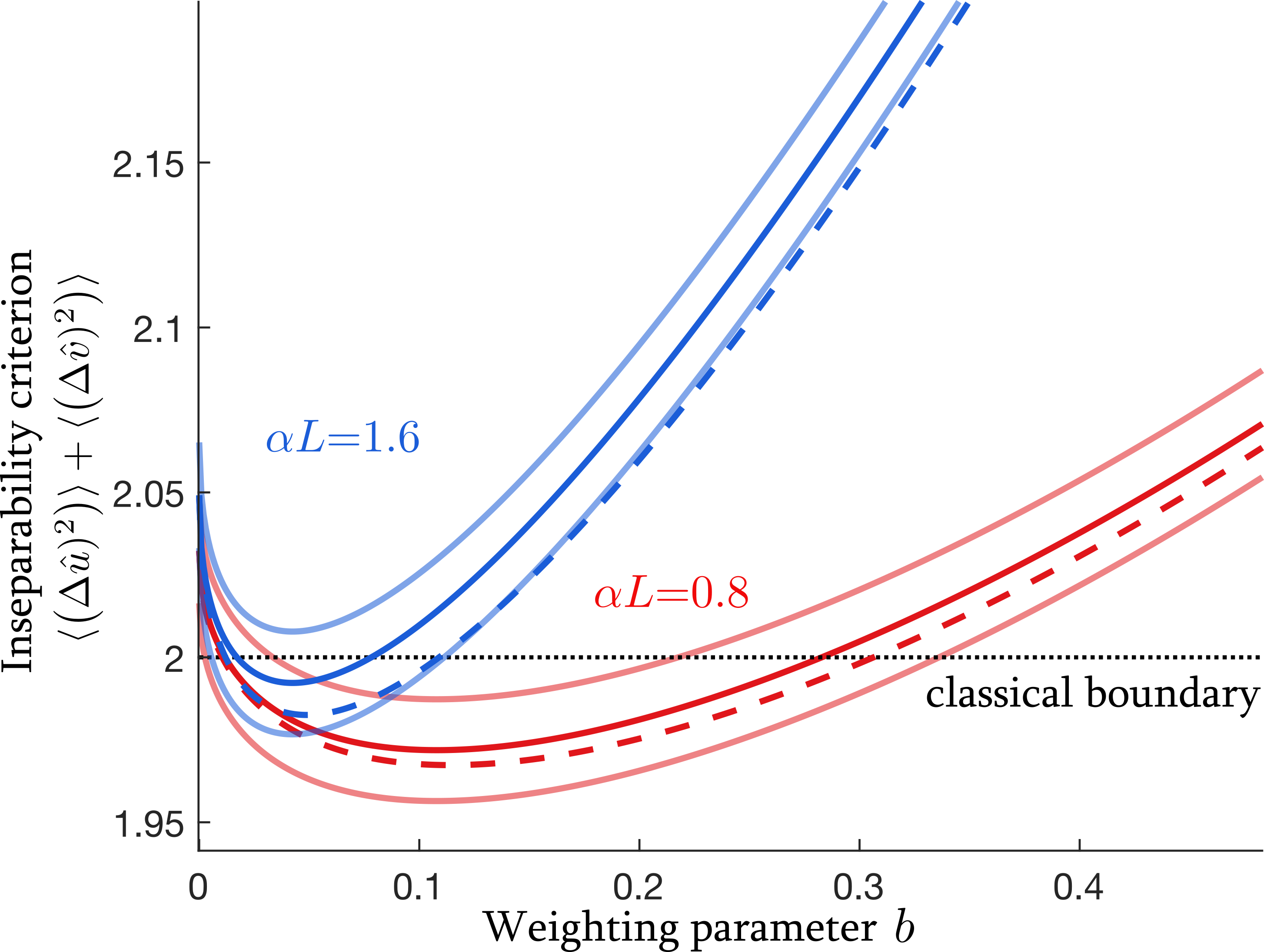}
    \caption{The inseparability criterion (see equation \ref{eq:insep}) as a function of weighting parameter (see equation \ref{eq:op}) for different optical depths $\alpha L$. Solid lines show the experimental result and 1$\sigma$ error averaging over 9,000 measurements using $\tau_{s}$ = \SI{5}{\micro\second}, dashed lines show the model.
  }  \label{fig:insep}
\end{figure}
Finally, we quantify the non-classical correlation between the amplified and rephased fields of the RASE experiment using the inseparability criterion \cite{Duan2000} at two optical depths (0.8 and 1.6). The criterion is constructed by expressing a maximally entangled state as a co-eigenstate of a pair of Einstein-Podolsky-Rosen (EPR)-type operators \cite{Einstein1935},
\begin{align} \label{eq:op}
\hat{u}&=\sqrt{b}\hat{x}_{A}+\sqrt{1-b}\hat{x}_{R},\\
\hat{v}&=\sqrt{b}\hat{p}_{A}-\sqrt{1-b}\hat{p}_{R},
\end{align}
where  $b\in$ [0,1] is an adjustable parameter describing the weight given to the ASE and RASE fields. For any separable state, the sum of the variance of the EPR-type operators satisfies
\begin{equation} \label{eq:insep}
\langle \left(\Delta\hat{u})^{2}\right)\rangle+ \langle \left(\Delta\hat{v})^{2}\right) \rangle \geq 2.
\end{equation}
For any inseparable state the total variance is bounded below by zero.  
Fig. \ref{fig:insep} shows the inseparability criterion as a function of the weighting parameter $b$ using a spin-storage time of \SI{5}{\micro\second}. When $b$ = 0 (1) the curve is purely the summation of the variance in the RASE (ASE) quadratures. The value of $b$ corresponding to the curve minimum suggests how the fields need to be weighted to maximise the correlation between the fields. At the lower optical depth (0.8), the curve minimum is 1.972(15) at $b$ = 0.11, violating the inseparability criterion with  1.8$\sigma$ confidence. At the higher optical depth (1.6), the curve minimum is 1.992(15) at $b$ = 0.04, violating the inseparability criterion with 0.5$\sigma$ confidence. In both cases, the non-classical result was maximised by reducing the detection windows from \SI{10}{\micro\second} to \SI{4}{\micro\second}.  

The theoretical inseparability criterion curves shown in Fig. \ref{fig:insep} were calculated using the model in Ref \cite{Ferguson2016}. In this model, the  ASE and RASE fields are assumed to initially be maximally entangled. The transmission losses (see above), that act on both fields equally, are then accounted for. Extra loss on the RASE field due to imperfect rephasing is also included. Each source of loss is assumed to only attenuate the field, mixing in a vacuum state.

%\section{\label{sec:Discussion} Discussion}
At an optical depth of $\alpha L$ = 0.8 there is good agreement between the observed and predicted inseparability curves. This indicates that there is little added classical noise or mixing in of states other than the vacuum state associated with the loss in the experiment. For $\alpha L$ = 1.6 the agreement between the model and the observations is marginal suggesting the level of added noise is becoming significant. This could be due to the classical noise, such as the phase noise, on the laser having a larger impact on the brighter fields, or could be noise directly related to the rephasing process. Both the loss mechanisms proposed in section \ref{sec:gain}, to explain the high level of ASE for $\alpha L >$ 2.0, are also potential sources of degradation of the correlation between the ASE and the RASE fields. Any population off-resonantly pumped into $|g_{2}\rangle$ will result in emission at the RASE frequency after the second rephasing pulse is applied which will be uncorrelated with the ASE field. Distortion of the spatial modes due to a spatially-dependent high optical depth will mix RASE fields corresponding to different ASE modes.

\section{\label{sec:Conclusion} Conclusion}
We have investigated the RASE efficiency as function of optical depth of the gain feature in Pr:\yso. We demonstrate that at an optical depth of $\alpha L$ = 0.8 the efficiency is maximized at roughly 14 \% and confirmed a non-classical correlation between the ASE and RASE fields. We also show that at optical depths of $\alpha L >$ 2.0 there are mechanisms in operation which reduce the correlation. To further improve the efficiency of the RASE protocol it will be necessary to increase the optical depth on the ASE and RASE transitions whilst maintaining or reducing the optical depth experienced by the control pulses. This could be achieved by employing an energy-level scheme in which the control fields excite transitions with lower oscillator strength, or through the use of an optical cavity, such that the ASE and RASE fields are supported by the cavity modes and the control pulses are applied in modes which are not supported. The use of an optical cavity could also reduce the noise associated with the mode distortion between the ASE and RASE fields by selectively enhancing the amplitude of the desired spatial modes.

\def\bibsection{

\end{document}
\section*{References}}

\begin{thebibliography}{}%

\bibitem{Kimble2008}
  \bibinfo{author}{H.~J. Kimble},
  \bibinfo{journal}{Nature}
  \textbf{\bibinfo{volume}{453}}, 
  \bibinfo{pages}{1023}
  (\bibinfo{year}{2008}).
  
\bibitem{Burnham1970}%
  \bibinfo{author}{D.~C. Burnham}\ and\,
  \bibinfo{author}{D.~L. Weinberg},
  \bibinfo{journal}{Physical Review Letters}
  \textbf{\bibinfo{volume}{25}},
  \bibinfo{pages}{84} 
  (\bibinfo{year}{1970}).
  
\bibitem{Fekete2013}%
  \bibinfo{author}{J. Fekete}, 
  \bibinfo{author}{D. Riel{\"a}nder}, 
  \bibinfo{author}{M. Cristiani}\ and\ 
  \bibinfo{author}{H. de Riedmatten},
  \bibinfo{journal}{Physical Review Letters}
  \textbf{\bibinfo {volume}{110}},
  \bibinfo{pages}{220502} 
  (\bibinfo {year}{2013}).
  
\bibitem{Duan2001}%
  \bibinfo{author}{L.-M. Duan},
  \bibinfo{author}{M.~D. Lukin},
  \bibinfo{author}{J.~I. Cirac}\ and\
  \bibinfo{author}{P. Zoller},
  \bibinfo {journal}{Nature}
  \textbf{\bibinfo{volume}{414}},
  \bibinfo{pages}{413}
  (\bibinfo{year}{2001}).
  
\bibitem{Ledingham2010}%
  \bibinfo{author}{P.~M. Ledingham},
  \bibinfo{author}{W.~R. Naylor},
  \bibinfo{author}{J.~J. Longdell}, 
  \bibinfo{author}{S.~E. Beavan}\ and\
  \bibinfo{author}{M.~J. Sellars},
  \bibinfo{journal}{Physical Review A}
  \textbf{\bibinfo{volume}{81}},
  \bibinfo {pages}{012301} 
  (\bibinfo {year} {2010}).
  
\bibitem{Hahn1950}%
  \bibinfo{author}{E.~L. Hahn},
  \bibinfo{journal}{Physical Review}
  \textbf{\bibinfo{volume}{80}},
  \bibinfo{pages}{580}
  (\bibinfo{year}{1950}).
  
\bibitem{Beavan2011}
  \bibinfo{author}{S.~E. Beavan},
  \bibinfo{author}{P.~M. Ledingham}, 
  \bibinfo{author}{J.~J. Longdell}\ and\
  \bibinfo{author}{M.~J. Sellars},
  \bibinfo{journal}{Optics Letters}
  \textbf{\bibinfo{volume}{36}},
  \bibinfo{pages}{1272}
  (\bibinfo{year}{2011}).
 

\bibitem{Ottaviani2009}%
  \bibinfo{author}{C. Ottaviani},
  \bibinfo{author}{C. Simon},
  \bibinfo{author}{H. de Riedmatten},
  \bibinfo{author}{M. Afzelius},
  \bibinfo{author}{B. Lauritzen},
  \bibinfo{author}{N. Sangouard}\ and\
  \bibinfo{author}{N. Gisin},
  \bibinfo{journal}{Physical Review A}
  \textbf{\bibinfo {volume}{79}},
  \bibinfo {pages}{063828}
  (\bibinfo{year}{2009}).
  
\bibitem{Ledingham2012}
  \bibinfo{author}{P.~M. Ledingham},
  \bibinfo{author}{W.~R. Naylor}\ and\
  \bibinfo{author}{J.~J. Longdell}, 
  \bibinfo{journal}{Physical Review Letters}
  \textbf{\bibinfo{volume}{109}},
  \bibinfo{pages}{093602}
  (\bibinfo{year}{2012}).
  
\bibitem{Beavan2012}
  \bibinfo{author}{S.~E. Beavan},
  \bibinfo{author}{M.~P. Hedges}\ and\
  \bibinfo{author}{M.~J. Sellars},
  \bibinfo{journal}{Physical Review Letters}
  \textbf{\bibinfo{volume}{109}},
  \bibinfo{pages}{093603}
  (\bibinfo{year}{2012}).
  
\bibitem{Ferguson2016} 
  \bibinfo{author}{K.~R. Ferguson},
  \bibinfo{author}{S.~E. Beavan},
  \bibinfo{author}{J.~J. Longdell}\ and\
  \bibinfo{author}{M.~J. Sellars},
  \bibinfo{journal}{Physical Review Letters}
  \textbf{\bibinfo{volume}{117}},
  \bibinfo{pages}{020501}
  (\bibinfo {year} {2016}).
  
\bibitem{Laplane2017}
  \bibinfo{author}{C. Laplane},
  \bibinfo{author}{P. Jobez},
  \bibinfo{author}{J. Etesse},
  \bibinfo{author}{N. Gisin}\ and\
  \bibinfo{author}{M. Afzelius},
  \bibinfo{journal}{Physical Review Letters}
  \textbf{\bibinfo{volume}{118}},
  \bibinfo{pages}{210501}
  (\bibinfo{year}{2017}).
  
\bibitem{Kutluer2017}
  \bibinfo{author}{K. Kutluer},
  \bibinfo{author}{M. Mazzera}\ and\
  \bibinfo{author}{H. de Riedmatten},
  \bibinfo {journal}{Physical Review Letters}
  \textbf{\bibinfo{volume}{118}},
  \bibinfo{pages}{210502}
  (\bibinfo {year} {2017}).
  
\bibitem{Stevenson2014}
  \bibinfo{author}{R.~N. Stevenson},
  \bibinfo{author}{M.~R. Hush},
  \bibinfo{author}{A.~R.~R. Carvalho},
  \bibinfo{author}{S.~E. Beavan}, 
  \bibinfo{author}{M.~J. Sellars}\ and\
  \bibinfo{author}{J.~J. Hope},
  \bibinfo{journal}{New Journal of Physics}
  \textbf{\bibinfo{volume}{16}},
  \bibinfo{pages}{033042}
  (\bibinfo{year}{2014}).
  
\bibitem{Jobez2014}
  \bibinfo{author}{P. Jobez},
  \bibinfo{author}{I. Usmani},
  \bibinfo{author}{N. Timoney},
  \bibinfo{author}{C. Laplane},
  \bibinfo {author}{N. Gisin}\ and\
  \bibinfo{author}{M. Afzelius},
  \bibinfo{journal}{New Journal of Physics}
  \textbf{\bibinfo{volume}{16}},
  \bibinfo{pages}{083005}
  (\bibinfo{year}{2014}).
  
\bibitem{Equall1995}
  \bibinfo{author}{R.~W. Equall},
  \bibinfo{author}{R.~L. Cone}\ and\
  \bibinfo{author}{R.~M. Macfarlane},
  \bibinfo{journal}{Physical Review B}
  \textbf{\bibinfo {volume}{52}},
  \bibinfo{pages}{3963}
  (\bibinfo{year}{1995}).
  
\bibitem{Nilsson2004}
  \bibinfo{author}{M. Nilsson},
  \bibinfo{author}{L. Rippe},
  \bibinfo{author}{S. Kroll},
  \bibinfo{author}{R. Klieber}\ and\
  \bibinfo{author}{D. Suter},
  \bibinfo{journal}{Physical Review B}
  \textbf{\bibinfo{volume}{70}},
  \bibinfo{pages}{214116}
  (\bibinfo{year}{2004}).
  
\bibitem{Duan2000}
  \bibinfo{author}{L.-M. Duan}, 
  \bibinfo{author}{G. Giedke},
  \bibinfo{author}{J.~I. Cirac}\ and\
  \bibinfo{author}{P. Zoller},
  \bibinfo{journal}{Physical Review Letters}
  \textbf{\bibinfo{volume}{84}},
  \bibinfo{pages}{2722}
  (\bibinfo{year}{2000}).
  
\bibitem{Einstein1935}
  \bibinfo{author}{A. Einstein},
  \bibinfo{author}{B. Podolsky}\ and\
  \bibinfo{author}{N. Rosen},
  \bibinfo{journal}{Physical Review}
  \textbf{\bibinfo{volume}{47}},
  \bibinfo{pages}{777}
  (\bibinfo{year}{1935}).
  
\end{thebibliography}
